\documentclass[11pt]{article}
\usepackage{hyperref}
\usepackage{latexsym,amsmath,amssymb,graphicx}
\usepackage{slashed, color}
\usepackage{multirow}
\usepackage{tikz}
\usepackage{amsmath}
\allowdisplaybreaks[2]
\usepackage{color}
\usepackage[multiple]{footmisc}
\usepackage{textcomp}
\usepackage{amssymb}
\usetikzlibrary{positioning}
\usepackage{amscd}
\usepackage{amsthm}
\usepackage{array} 

\renewcommand{\thefootnote}{\fnsymbol{footnote}}
\numberwithin{equation}{section}
\usepackage{dsfont}
\usepackage{mathrsfs} 
\usepackage{booktabs}

\DeclareFontFamily{U}{MnSymbolC}{}
\DeclareSymbolFont{MnSyC}{U}{MnSymbolC}{m}{n}
\DeclareFontShape{U}{MnSymbolC}{m}{n}{
    <-6>  MnSymbolC5
   <6-7>  MnSymbolC6
   <7-8>  MnSymbolC7
   <8-9>  MnSymbolC8
   <9-10> MnSymbolC9
  <10-12> MnSymbolC10
  <12->   MnSymbolC12}{}
\DeclareMathSymbol{\intprod}{\mathbin}{MnSyC}{'270}
\newcommand{\ov}{\overline}
\newcommand{\C}{\mathbb{C}}

\newcommand{\Z}{\mathbb{Z}}
\newcommand{\R}{\mathbb{R}}

\newcommand{\del}{\partial}

\newcommand{\til}{\widetilde}

\let\nc\newcommand
\let\renc\renewcommand
\nc{\wbar}{\overline}
\let\td\tilde
\let\wtd\widetilde
\let\wht\widehat
\let\mcl\mathcal

\nc{\ab}{{\bar{a}}} \nc{\at}{\tilde{a}} \nc{\ah}{\hat{a}}
\nc{\bb}{{\bar{b}}} 
\nc{\bh}{\hat{b}}
\nc{\cb}{{\bar{c}}} \nc{\ct}{\tilde{c}} 
\nc{\db}{{\bar{d}}} \nc{\dt}{\tilde{d}} \renc{\dh}{\hat{d}}
\nc{\eb}{{\bar{e}}} \nc{\et}{\tilde{e}} \nc{\eh}{\hat{e}}
\nc{\fb}{{\bar{f}}} \nc{\ft}{\tilde{f}} \nc{\fh}{\hat{f}}
\nc{\gb}{{\bar{g}}} \nc{\gt}{\tilde{g}} \nc{\gh}{\hat{g}}
\nc{\ib}{{\bar{\imath}}} \nc{\ih}{\hat{\imath}} 
\nc{\jb}{{\bar{\jmath}}} \nc{\jt}{\tilde{\jmath}} \nc{\jh}{\hat{\jmath}}
\nc{\kb}{{\bar{k}}} \nc{\kt}{\tilde{k}} \nc{\kh}{\hat{k}}
\nc{\lb}{{\bar{l}}} \nc{\lt}{\tilde{l}} \nc{\lh}{\hat{l}}
\nc{\mb}{{\bar{m}}} \nc{\mt}{\tilde{m}} \nc{\mh}{\hat{m}}
\nc{\nb}{{\bar{n}}} \nc{\nt}{\tilde{n}} \nc{\nh}{\hat{n}}
\nc{\ob}{{\bar{o}}} \nc{\ot}{\tilde{o}} \nc{\oh}{\hat{o}}
\nc{\pb}{{\bar{p}}} \nc{\pt}{\tilde{p}} \nc{\ph}{\hat{p}}
\nc{\qb}{{\bar{q}}} \nc{\qt}{\tilde{q}} \nc{\qh}{\hat{q}}
\nc{\rb}{{\bar{r}}} \nc{\rt}{\tilde{r}} \nc{\rh}{\hat{r}}
\renc{\sb}{{\bar{s}}} \nc{\st}{\tilde{s}} \nc{\sh}{\hat{s}}
\nc{\tb}{{\bar{t}}} \renc{\th}{\hat{t}} 
\nc{\ub}{{\bar{u}}} \nc{\ut}{\tilde{u}} \nc{\uh}{\hat{u}}
\nc{\vb}{{\bar{v}}} \nc{\vt}{\tilde{v}} \nc{\vh}{\hat{v}}
\nc{\wt}{\tilde{w}} \nc{\wh}{\hat{w}}
\nc{\xb}{{\bar{x}}} \nc{\xt}{\tilde{x}} \nc{\xh}{\hat{x}}
\nc{\yb}{{\bar{y}}} \nc{\yt}{\tilde{y}} \nc{\yh}{\hat{y}}
\nc{\zb}{{\bar{z}}} \nc{\zt}{\tilde{z}} 

\nc{\Ab}{\wbar{A}} \nc{\At}{\wtd{A}} \nc{\Ah}{\wht{A}}
\nc{\Bb}{\wbar{B}} \nc{\Bt}{\wtd{B}} \nc{\Bh}{\wht{B}}
\nc{\Cb}{\wbar{C}} \nc{\Ct}{\wtd{C}} \nc{\Ch}{\wht{C}}
\nc{\Db}{\wbar{D}} \nc{\Dt}{\wtd{D}} \nc{\Dh}{\wht{D}}
\nc{\Eb}{\wbar{E}} \nc{\Et}{\wtd{E}} \nc{\Eh}{\wht{E}}
\nc{\Fb}{\wbar{F}} \nc{\Ft}{\wtd{F}} \nc{\Fh}{\wht{F}}
\nc{\Gb}{\wbar{G}} \nc{\Gt}{\wtd{G}} \nc{\Gh}{\wht{G}}
\nc{\Hb}{\wbar{H}} \nc{\Ht}{\wtd{H}} \nc{\Hh}{\wht{H}}
\nc{\Ib}{\wbar{I}} \nc{\It}{\wtd{I}} \nc{\Ih}{\wht{I}}
\nc{\Jb}{\wbar{J}} \nc{\Jt}{\wtd{J}} \nc{\Jh}{\wht{J}}
\nc{\Kb}{\wbar{K}} \nc{\Kt}{\wtd{K}} \nc{\Kh}{\wht{K}}
\nc{\Lb}{\wbar{L}} \nc{\Lt}{\wtd{L}} \nc{\Lh}{\wht{L}}
\nc{\Mb}{\wbar{M}} \nc{\Mt}{\wtd{M}} \nc{\Mh}{\wht{M}}
\nc{\Nb}{\wbar{N}} \nc{\Nt}{\wtd{N}} \nc{\Nh}{\wht{N}}
\nc{\Ob}{\wbar{O}} \nc{\Ot}{\wtd{O}} \nc{\Oh}{\wht{O}}
\nc{\Pb}{\wbar{P}} \nc{\Pt}{\wtd{P}} \nc{\Ph}{\wht{P}}
\nc{\Qb}{\wbar{Q}} \nc{\Qt}{\wtd{Q}} \nc{\Qh}{\wht{Q}}
\nc{\Rb}{\wbar{R}} \nc{\Rt}{\wtd{R}} \nc{\Rh}{\wht{R}}
\nc{\Sb}{\wbar{S}} \nc{\St}{\wtd{S}} \nc{\Sh}{\wht{S}}
\nc{\Tb}{\wbar{T}} \nc{\Tt}{\wtd{T}} \nc{\Th}{\wht{T}}
\nc{\Ub}{\wbar{U}} \nc{\Ut}{\wtd{U}} \nc{\Uh}{\wht{U}}
\nc{\Vb}{\wbar{V}} \nc{\Vt}{\wtd{V}} \nc{\Vh}{\wht{V}}
\nc{\Wb}{\wbar{W}} \nc{\Wt}{\wtd{W}} \nc{\Wh}{\wht{W}}
\nc{\Xb}{\wbar{X}} \nc{\Xt}{\wtd{X}} \nc{\Xh}{\wht{X}}
\nc{\Yb}{\wbar{Y}} \nc{\Yt}{\wtd{Y}} \nc{\Yh}{\wht{Y}}
\nc{\Zb}{\wbar{Z}} \nc{\Zt}{\wtd{Z}} \nc{\Zh}{\wht{Z}}

\nc{\CA}{\mcl{A}} \nc{\CAb}{\wbar{\CA}} \nc{\CAt}{\wtd{\CA}} \nc{\CAh}{\wht{\CA}}
\nc{\CB}{\mcl{B}} \nc{\CBb}{\wbar{\CB}} \nc{\CBt}{\wtd{\CB}} \nc{\CBh}{\wht{\CB}}
\nc{\CC}{\mcl{C}} \nc{\CCb}{\wbar{\CC}} \nc{\CCt}{\wtd{\CC}} \nc{\CCh}{\wht{\CC}}
\nc{\cDt}{\wtd{\cC}} \nc{\cDh}{\wht{\cD}}
\nc{\CE}{\mcl{E}} \nc{\CEb}{\wbar{\CE}} \nc{\CEt}{\wtd{\CE}} \nc{\CEh}{\wht{\CE}}
\nc{\CF}{\mcl{F}} \nc{\CFb}{\wbar{\CF}} \nc{\CFt}{\wtd{\CF}} \nc{\CFh}{\wht{\CF}}
\nc{\CG}{\mcl{G}} \nc{\CGb}{\wbar{\CG}} \nc{\CGt}{\wtd{\CG}} \nc{\CGh}{\wht{\CG}}
\nc{\CH}{\mcl{H}} \nc{\CHb}{\wbar{\CH}} \nc{\CHt}{\wtd{\CH}} \nc{\CHh}{\wht{\CH}}
\nc{\CI}{\mcl{I}} \nc{\CIb}{\wbar{\CI}} \nc{\CIt}{\wtd{\CI}} \nc{\CIh}{\wht{\CI}}
\nc{\CJ}{\mcl{J}} \nc{\CJb}{\wbar{\CJ}} \nc{\CJt}{\wtd{\CJ}} \nc{\CJh}{\wht{\CJ}}
\nc{\CK}{\mcl{K}} \nc{\CKb}{\wbar{\CK}} \nc{\CKt}{\wtd{\CK}} \nc{\CKh}{\wht{\CK}}
\nc{\CL}{\mcl{L}} \nc{\CLb}{\wbar{\CL}} \nc{\CLt}{\wtd{\CL}} \nc{\CLh}{\wht{\CL}}
\nc{\CM}{\mcl{M}} \nc{\CMb}{\wbar{\CM}} \nc{\CMt}{\wtd{\CM}} \nc{\CMh}{\wht{\CM}}
\nc{\CN}{\mcl{N}} \nc{\CNb}{\wbar{\CN}} \nc{\CNt}{\wtd{\CN}} \nc{\CNh}{\wht{\CN}}
\nc{\CO}{\mcl{O}} \nc{\COb}{\wbar{\CO}} \nc{\COt}{\wtd{\CO}} \nc{\COh}{\wht{\CO}}
\nc{\CQ}{\mcl{Q}} \nc{\CQb}{\wbar{\CQ}} \nc{\CQt}{\wtd{\CQ}} \nc{\CQh}{\wht{\CQ}}
\nc{\CR}{\mcl{R}} \nc{\CRb}{\wbar{\CR}} \nc{\CRt}{\wtd{\CR}} \nc{\CRh}{\wht{\CR}}
\nc{\CS}{\mcl{S}} \nc{\CSb}{\wbar{\CS}} \nc{\CSt}{\wtd{\CS}} \nc{\CSh}{\wht{\CS}}
\nc{\CT}{\mcl{T}} \nc{\CTb}{\wbar{\CT}} \nc{\CTt}{\wtd{\CT}} \nc{\CTh}{\wht{\CT}}
\nc{\CU}{\mcl{U}} \nc{\CUb}{\wbar{\CU}} \nc{\CUt}{\wtd{\CU}} \nc{\CUh}{\wht{\CU}}
\nc{\CV}{\mcl{V}} \nc{\CVb}{\wbar{\CV}} \nc{\CVt}{\wtd{\CV}} \nc{\CVh}{\wht{\CV}}
\nc{\CW}{\mcl{W}} \nc{\CWb}{\wbar{\CW}} \nc{\CWt}{\wtd{\CW}} \nc{\CWh}{\wht{\CW}}
\nc{\CX}{\mcl{X}} \nc{\CXb}{\wbar{\CX}} \nc{\CXt}{\wtd{\CX}} \nc{\CXh}{\wht{\CX}}
\nc{\CY}{\mcl{Y}} \nc{\CYb}{\wbar{\CY}} \nc{\CYt}{\wtd{\CY}} \nc{\CYh}{\wht{\CY}}
\nc{\CZ}{\mcl{Z}} \nc{\CZb}{\wbar{\CZ}} \nc{\CZt}{\wtd{\CZ}} \nc{\CZh}{\wht{\CZ}}

\let\eps\epsilon
\let\ups\upsilon
\let\veps\varepsilon
\let\vtht\vartheta
\let\vsgm\varsigma
\let\vphi\varphi
\let\vrho\varrho

\nc{\alphab}{\bar{\alpha}} \nc{\alphat}{\td{\alpha}} \nc{\alphah}{\hat{\alpha}}
\nc{\betab}{\bar{\beta}}   \nc{\betat}{\td{\beta}}   \nc{\betah}{\hat{\beta}} 
\nc{\gammab}{\bar{\gamma}} \nc{\gammat}{\td{\gamma}} \nc{\gammah}{\hat{\gamma}} 
\nc{\deltab}{\bar{\delta}} \nc{\deltat}{\td{\delta}} \nc{\deltah}{\hat{\delta}} 
\nc{\epsilonb}{\bar{\eps}} \nc{\epsilont}{\td{\eps}} \nc{\epsilonh}{\hat{\eps}} 
\nc{\vepsb}{\bar{\veps}}   \nc{\vepst}{\td{\veps}}   \nc{\vepsh}{\hat{\veps}} 
\nc{\zetab}{\bar{\zeta}}   \nc{\zetat}{\td{\zeta}}   \nc{\zetah}{\hat{\zeta}} 
\nc{\etab}{\bar{\eta}}     
\nc{\etah}{\hat{\eta}} 
\nc{\thetab}{\bar{\theta}} \nc{\thetat}{\td{\theta}} \nc{\thetah}{\hat{\theta}} 
\nc{\vthetab}{\bar{\vtht}} \nc{\vthetat}{\td{\vtht}} \nc{\vthetah}{\hat{\vtht}} 
\nc{\lambdat}{\td{\lambda}} \nc{\lambdah}{\hat{\lambda}} 
\nc{\iotab}{\bar{\iota}}   \nc{\iotat}{\td{\iota}}   \nc{\iotah}{\hat{\iota}} 
\nc{\kappab}{\bar{\kappa}} \nc{\kappat}{\td{\kappa}} \nc{\kappah}{\hat{\kappa}} 
\nc{\lmdb}{\bar{\lmd}}     \nc{\lmdt}{\td{\lmd}}     \nc{\lmdh}{\hat{\lmd}} 
\nc{\mub}{\bar{\mu}}       \nc{\mut}{\td{\mu}}       \nc{\muh}{\hat{\mu}} 
\nc{\nub}{\bar{\nu}}       \nc{\nut}{\td{\nu}}       \nc{\nuh}{\hat{\nu}} 
\nc{\xib}{\bar{\xi}}       \nc{\xit}{\td{\xi}}       \nc{\xih}{\hat{\xi}} 
\nc{\pib}{\bar{\pi}}       \nc{\pit}{\td{\pi}}       \nc{\pih}{\hat{\pi}} 
\nc{\vpib}{\bar{\vpi}}     \nc{\vpit}{\td{\vpi}}     \nc{\vpih}{\hat{\vpi}} 
\nc{\rhob}{\bar{\rho}}     \nc{\rhot}{\td{\rho}}     \nc{\rhoh}{\hat{\rho}} 
\nc{\vrhob}{\bar{\vrho}}   \nc{\vrhot}{\td{\vrho}}   \nc{\vrhoh}{\hat{\vrho}} 
\nc{\sigmab}{\bar{\sigma}} \nc{\sigmat}{\td{\sigma}} \nc{\sigmah}{\hat{\sigma}} 
\nc{\vsigmab}{\bar{\vsgm}} \nc{\vsigmat}{\td{\vsgm}} \nc{\vsigmah}{\hat{\vsgm}} 
\nc{\taub}{\bar{\tau}}     \nc{\taut}{\td{\tau}}     \nc{\tauh}{\hat{\tau}} 
\nc{\upsb}{\bar{\ups}} \nc{\upst}{\td{\ups}} \nc{\upsh}{\hat{\ups}} 
\nc{\phib}{\bar{\phi}}     \nc{\phit}{\td{\phi}}     \nc{\phih}{\hat{\phi}} 
\nc{\varphib}{\bar{\vphi}}   \nc{\varphit}{\td{\vphi}}   \nc{\varphih}{\hat{\vphi}} 
\nc{\chib}{\bar{\chi}}     
\nc{\chih}{\hat{\chi}} 
\nc{\psib}{\bar{\psi}}     
\nc{\psih}{\hat{\psi}} 
\nc{\omegab}{\bar{\omega}} \nc{\omegat}{\td{\omega}} \nc{\omegah}{\hat{\omega}} 

\nc{\Gammab}{\wbar{\Gamma}}     \nc{\Gammat}{\wtd{\Gamma}}     \nc{\Gammah}{\wht{\Gamma}}
\nc{\Deltab}{\wbar{\Delta}}     \nc{\Deltat}{\wtd{\Delta}}     \nc{\Deltah}{\wht{\Delta}}
\nc{\Thetab}{\wbar{\Theta}}     \nc{\Thetat}{\wtd{\Theta}}     \nc{\Thetah}{\wht{\Theta}}
\nc{\Lambdab}{\wbar{\Lambda}}   \nc{\Lambdat}{\wtd{\Lambda}}   \nc{\Lambdah}{\wht{\Lambda}}
\nc{\Xib}{\wbar{\Xi}}           \nc{\Xit}{\wtd{\Xi}}           \nc{\Xih}{\wht{\Xi}}
\nc{\Pib}{\wbar{\Pi}}           \nc{\Pit}{\wtd{\Pi}}           \nc{\Pih}{\wht{\Pi}}
\nc{\Sigmab}{\wbar{\Sigma}}     \nc{\Sigmat}{\wtd{\Sigma}}     \nc{\Sigmah}{\wht{\Sigma}}
\nc{\Upsilonb}{\wbar{\Upsilon}} \nc{\Upsilont}{\wtd{\Upsilon}} \nc{\Upsilonh}{\wht{\Upsilon}}
\nc{\Phib}{\wbar{\Phi}}         \nc{\Phit}{\wtd{\Phi}}         \nc{\Phih}{\wht{\Phi}}
\nc{\Psib}{\wbar{\Psi}}         \nc{\Psit}{\wtd{\Psi}}         \nc{\Psih}{\wht{\Psi}}
\nc{\Omegab}{\wbar{\Omega}}     \nc{\Omegat}{\wtd{\Omega}}     \nc{\Omegah}{\wht{\Omega}}

\newcommand{\cA}{{\cal A}}
\newcommand{\cAb}{{\overline{\cal A}}}

\def\bear{\begin{eqnarray}}
\def\eear{\end{eqnarray}}

\newcommand{\Tr}{{\textrm{Tr}\;}}

\newcommand{\cF}{{\cal F}}
\newcommand{\cFb}{{\overline{\cal F}}}

\newcommand{\quar}{\frac{1}{4}}

\newcommand{\cD}{{\cal D}}
\newcommand{\cDb}{{\overline{\cal D}}}

\newcommand{\hlf}{\frac{1}{2}}

\newcommand{\mA}{\mathcal{A}}

\newcommand{\si}{\sigma}

\newcommand{\hM}{\widehat{M}}
\newcommand{\hN}{\widehat{N}}
\newcommand{\hm}{\widehat{m}}
\newcommand{\hn}{\widehat{n}}
\newcommand{\ha}{\widehat{A}}
\newcommand{\hb}{\widehat{B}}
\newcommand{\hc}{\widehat{C}}
\newcommand{\x}{$\times$}
\nc{\balpha}{\bar{\alpha}}
\nc{\bbeta}{\bar{\beta}}
\nc{\bgamma}{\bar{\gamma}}
\nc{\bm}{\bar{m}}
\nc{\bn}{\bar{n}}
\nc{\bp}{\bar{p}}
\nc{\al}{\alpha}
\nc{\bt}{\beta}
\nc{\gm}{\gamma}
\nc{\zh}{\wht{z}}
\nc{\zhb}{\ov{\wht{z}}}
\nc{\mbh}{\wht{\ov{m}}}
\nc{\bc}{|_{x^3=0}}

\nc{\tal}{\til{\al}}
\nc{\tbt}{\til{\bt}}
\nc{\tgm}{\til{\gm}}

\nc{\wb}{\ov{w}}
\nc{\teta}{\til{\eta}}
\nc{\tpsi}{\til{\psi}}

\let\OLDthebibliography\thebibliography
\renewcommand\thebibliography[1]{
  \OLDthebibliography{#1}
  \setlength{\parskip}{5pt}
  \setlength{\itemsep}{0pt plus 0.3ex}
}
\usepackage{titlesec}
\titleformat*{\section}{\bfseries\large}
\begin{document}
\addtolength{\baselineskip}{1.5mm}

\thispagestyle{empty}
\vbox{}
\vspace{3.0cm}

\begin{center}
\centerline{\LARGE{Branes and Categorifying Integrable Lattice Models}}

\vspace{3.0cm}

{Meer~Ashwinkumar\footnote{E-mail: meerashwinkumar@u.nus.edu}, Meng-Chwan~Tan\footnote{E-mail: mctan@nus.edu.sg}, and Qin~Zhao\footnote{E-mail: zhaoqin@u.nus.edu}}
\\[2mm]
{\it Department of Physics\\
National University of Singapore \\
2 Science Drive 3, Singapore 117551} \\[1mm] 
\end{center}

\vspace{2.0cm}

\centerline{\bf Abstract}\smallskip \noindent
We elucidate how integrable lattice models described by Costello's 4d Chern-Simons theory can be realized via a stack of D4-branes ending on an NS5-brane in type IIA string theory, with D0-branes on the D4-brane worldvolume sourcing a meromorphic RR 1-form, and fundamental strings forming the lattice. This provides us with a nonperturbative integration cycle for the 4d Chern-Simons theory, 
and by applying T- and S-duality, we show how the R-matrix, the Yang-Baxter equation and the Yangian can be categorified, that is, obtained via the Hilbert space of a 6d gauge theory.

\newpage

\renewcommand{\thefootnote}{\arabic{footnote}}
\setcounter{footnote}{0}

\tableofcontents

\section{Introduction and summary}

The Yang-Baxter equation with spectral parameter was recently found to arise from a 4d variant of Chern-Simons gauge theory devised by Costello \cite{Costello,CWY,CWY2}, with the action 
\begin{equation}\label{introaction}
S=\frac{1}{\hbar}\int_{ Y\times \Sigma} C\wedge \textrm{Tr}\bigg(\cA\wedge d\cA + \frac{2}{3} \cA\wedge \cA\wedge \cA\bigg),
\end{equation}
where $\cA$ is a complex-valued gauge field, $Y$ is a framed 2-manifold, and $\Sigma$ is a complex Riemann surface endowed with a meromorphic one-form $C=C(z)dz$, which can have poles but no zeros. 
 Within the realm of perturbation theory, this gauge theory 
 encapsulates the underlying structure of integrable lattice models of two-dimensional classical statistical mechanics. 


Outside of perturbation theory, the 4d Chern-Simons theory \eqref{introaction} is not well-understood (its path integral is exponentially divergent), and it was suggested in \cite{Witten} that a nonperturbative definition of the theory 
should arise from the D4-NS5 brane system of type IIA string theory, in a manner similar to how analytically-continued 3d Chern-Simons theory can be given a nonperturbative definition via the D3-NS5 system. 
Our aim in this work is to firstly 
verify the suggestion in \cite{Witten}, and to derive the integration cycle 
which allows \eqref{introaction} to be well-defined beyond perturbation theory; this shall be done in Section 2. 

Secondly, given that the D3-NS5 brane embedding of 3d Chern-Simons theory leads to the categorification of knot polynomials in terms of Khovanov homology \cite{WittenKho}, it is natural to ask if the D4-NS5 brane system will lead us to a categorification of the Yang-Baxter equation. Indeed, in Section 3, by applying T- and S-duality to arrive at the NS5-D5 system in type IIB string theory, we will categorify the elements of the R-matrix which solves the aforementioned Yang-Baxter equation, thereby categorifying the Yang-Baxter equation itself, and we shall furthermore categorify the Yangian. 

A brief summary of our results is as follows. We shall first show that a twisted sector of the D4-NS5 system with a meromorphic RR 1-form is equivalent to Costello's 4d Chern-Simons theory \eqref{box1} with nonperturbative integration cycle defined by \eqref{box2}. Using T- and S-duality, we arrive at the NS5-D5 brane system, the supersymmetric Hilbert space of ground states of which is defined by the Floer cohomology of the 6d equations \eqref{6dbps} which interpolate solutions of the 5d equations \eqref{5dbps}. Via these dualities, we are able to express each R-matrix element in terms of a trace over this Hilbert space in \eqref{boxR} (thereby categorifying each R-matrix element), and we are also able to categorify the Yang-Baxter equation, as shown in \eqref{boxYB}, as well as the Yangian.
\section{4d Chern-Simons theory from 5d topologically twisted MSYM coupled to meromorphic RR 1-form}
\subsection{D4-brane worldvolume theory with NS5 boundary conditions}
The low energy worldvolume theory of $N$ coincident D4-branes on a flat manifold, $M$, 
has the classical action
\cite{Geyer,CJ}
\begin{equation}
\begin{aligned}
\label{untwisted}
S = -\frac{1}{g_5^2} \int_M d^5x ~\Tr \Big(&\quar F_{MN} F^{MN} + \hlf D_M \phi_{\hM} D^M \phi^{\hM} + \quar [\phi_{\hM}, \phi_{\hN}] [\phi^{\hM}, \phi^{\hN}] 
+i{\rho}^{A{\ha}}(\Gamma^M)_A^{~~B} D_M \rho_{B\ha} \\&+ {\rho}^{A\ha} (\Gamma^{\hM})_{\ha}^{~~\hb} [\phi_{\hM}, \rho_{A\hb}] \Big),
\end{aligned}
\end{equation} 
i.e., 5d maximally supersymmetric Yang-Mills theory (MSYM), which is invariant under the supersymmetry transformations 
\begin{equation}\label{untwistedsusy}
\begin{aligned}
\delta A_M &= 2\zeta ^{A\ha} (\Gamma_M)_A^{~~B}\rho_{B\ha}\\
\delta \phi ^{\hM}&=-i2 \zeta ^{A\ha} (\Gamma^{\hM})_{\ha}^{~~\hb}\rho_{A\hb}\\
\delta\rho _{A\ha}&=(\Gamma^M)_A^{~~B}D_M\phi^{\hM}(\Gamma_{\hM})_{\ha}^{~~\hb}\zeta_{B\hb}-\frac{i}{2}(\Gamma_{\hM})_{\ha}^{~~\hb}(\Gamma_{\hN})_{{\hb}\hc}[\phi^{\hM},\phi^{\hN}]\zeta_A^{~~\hc}-\frac{i}{2}F^{MN}(\Gamma_{MN})_{AB}\zeta^B_{~{\ha}}.
\end{aligned}
\end{equation}
Here, $(M,N,\ldots)$ and $(A,B,\ldots)$ are respectively vector and spinor indices for the $SO_{E}(5)$ rotation group, with their hatted counterparts corresponding to the $SO_R(5)$ R-symmetry group. In addition, the Lie algebra of the $U(N)$ gauge group is taken to be generated by antihermitian matrices $T_a$, where $a=1,\ldots,\textrm{dim }\mathfrak{u}(N)$, implying that the invariant quadratic form on this Lie algebra, denoted $\textrm{Tr}$, is negative-definite. In particular, the matrices $T_a$ are chosen such that $\textrm{Tr}(T_a T_b)=-\delta_{ab}$.

We shall take the D4-branes to end on an NS5-brane, in the type IIA brane configuration in flat Euclidean space given by the following table:
\begin{center}
\begin{tabular}{@{}lllllllllll@{}}
\toprule
   & \textbf{1} & \textbf{2} & \textbf{3} & \textbf{4} & \textbf{5} & \textbf{6} & \textbf{7} & \textbf{8} & \textbf{9} & \textbf{10}\\
   \textbf{D4} & \x & \x & \x & \x & \x & & & & &\\
   \textbf{NS5} & \x & \x &  & \x & \x & \x & \x & & & \\
\bottomrule
 \end{tabular}
\end{center}
where, e.g., an empty entry under `3' indicates that the brane is located at $x^3=0$. Also note that the scalar fields $\{\phi_{\wht{1}},\phi_{\wht{2}},\phi_{\wht{3}},\phi_{\wht{4}},\phi_{\wht{5}}\}$ of the worldvolume theory are taken to parametrize the $\{6,7,8,9,10\}$ directions, respectively.
This configuration induces the following boundary conditions on the fields of the worldvolume theory:  
\begin{equation}\label{boundaryc}
\begin{aligned}
F_{\mu 3}=0|_{x^3=0}, && D_{3}\phi^{\wht{1},\wht{2}}=0|_{x^3=0}, && \phi^{\wht{3},\wht{4},\wht{5}}=0|_{x^3=0},
\end{aligned}
\end{equation}
(here, $\mu$ is the boundary $SO_E(4)$ rotation group 4d index) together with projection conditions on the fermionic fields.

We now wish to perform a partial topological twist of the worldvolume theory along a submanifold of $M$. The reason for this is we wish to obtain Costello's 4d Chern-Simons theory, which is a topological-holomorphic theory, and hence not fully topological. Let the flat manifold $M=Y\times \R_+ \times \Sigma$, where $Y$ and $\Sigma$ are 2-manifolds corresponding to the $\{x^1,x^2\}$ and $\{x^4,x^5\}$ directions respectively, while $\R_+$ is half of the real line, $\R$, that corresponds to the $x^3$ direction. We shall twist along $V=Y\times \R_+$, by redefining its $SO_E(3)$ rotation group to be the diagonal subgroup $SO_E(3)'$ of $SO_E(3)\times SO_R(3)$, where $SO_R(3)$ is the subgroup of the R-symmetry group that rotates $\{\phi_{\wht{1}},\phi_{\wht{2}},\phi_{\wht{3}}\}$. 

From the string theoretical perspective, this partial twist amounts to taking the $\{x^1,x^2,x^3\}$ and $\{x^6,x^7,x^8\}$ directions to form $T^*\til{V}$, where $\til{V}=Y\times \R$; the twist follows essentially because $V\subset\til{V}$ and $\til{V}$ is the zero section of the cotangent bundle $T^*\til{V}$, and therefore `coordinates' normal to $\til{V}$ in $T^*\til{V}$ must be components of one-forms \cite{Bershadsky}, as we shall obtain via the twisting procedure.

Let us denote $SO_E(3)$ vector indices by $(\alpha,\beta,\ldots)$ and  $SO_E(3)$ spinor indices by $(\bar{\alpha},\bar{\beta},\ldots)$, with their hatted versions corresponding to $SO_R(3)$. Twisting simply amounts to setting the hatted indices to unhatted indices.
The remaining $SO_E(2)$ vector indices not involved in twisting shall be denoted $(m,n,\ldots)$, with the barred and hatted versions having the usual meaning of spinor and R-symmetry indices. 

As a result of the twisting, the scalar fields $\{\phi_{\wht{1}},\phi_{\wht{2}},\phi_{\wht{3}}\}$ now transform as the components $\{\phi_{{1}},\phi_{{2}},\phi_{{3}}\}$ of a one-form on $Y\times \R_+$. In addition, the twisting of the fermions which transform as (\textbf{2},\textbf{2}) under $SO_E(3)\times SO_R(3)$ results in fermions which transform as \textbf{1} and \textbf{3}  under $SO_E(3)'$, i.e.,
\begin{equation}
\textbf{2} \otimes \textbf{2} = \textbf{1} \oplus \textbf{3}.
\end{equation}
To see this explicitly, the spinor fields $\rho_{A\wht{A}}=\rho_{\balpha \bm \wht{\balpha}\wht{\bm}}$ can be expanded after twisting as 
\begin{equation}\label{fermiexp}
\rho_{\balpha \bm {\bbeta}\wht{\bm}}= \eps_{\balpha \bbeta} \eta_{\bm \wht{\bm}} + (\si^\alpha)_{\balpha \bbeta} \psi_{\alpha\bm \wht{\bm}},
\end{equation}
where we have 
used the antisymmetric matrix $\epsilon_{\balpha \bbeta}$ and the symmetric matrix $(\si^\alpha)_{\balpha \bbeta}$ introduced in the appendix. The supersymmetry transformation parameters $\zeta_{A\wht{A}}=\zeta_{\balpha \bm \wht{\balpha}\wht{\bm}}$ can also be expanded in such a manner, i.e., 
\begin{equation}\label{susyexp}
\zeta_{\balpha \bm {\bbeta}\wht{\bm}}= \eps_{\balpha \bbeta} \zeta_{\bm \wht{\bm}} + (\si^\alpha)_{\balpha \bbeta} \zeta_{\alpha\bm \wht{\bm}}.
\end{equation}
With the explicit representation of the gamma matrices given in the appendix, we can substitute \eqref{fermiexp} and \eqref{susyexp} into \eqref{untwisted} and \eqref{untwistedsusy} to obtain the partially twisted action and supersymmetry transformations.

Now, we shall pick a particular supercharge, $Q$, that is scalar along $V$, with respect to which we shall eventually localize the theory. The supersymmetry transformations generated by this supercharge ought to leave invariant the combinations $A_1+i\phi_1$ (or $A_1-i\phi_1$), $A_2+i\phi_2$ (or $A_2-i\phi_2$), and $A_{\zb}=\frac{1}{2}(A_4+iA_5)$, since these combinations are the natural candidates for 
the fields of the 4d Chern-Simons theory, to which we hope to localize the partially twisted theory.
Indeed, there are two such supercharges, corresponding to the supersymmetry transformation parameters $\zeta_{11}$ and $\zeta_{12}$ in \eqref{susyexp}. Without any loss of generality, we shall pick the supercharge corresponding to $\zeta_{11}$. Observe that from the on-shell supersymmetry algebra of \eqref{untwisted},
\begin{equation}
\{Q_{A\hb},Q^{B\hc} \}=(\Gamma^M)_A^{~~B}\delta_{\hb}^{~~\hc}P_M-i\delta_A^{~~B}(\Gamma^{\hM})_{\hb}^{~~\hc}P_{\hM},
\end{equation}
where $P_M$ are the worldvolume momenta and $P_{\hM}$ are central charges, we find (by expanding $Q_{A\hb}$ as in \eqref{fermiexp} and \eqref{susyexp}) 
that the supercharge $Q_{22}$ which corresponds to $\zeta_{11}$ satisfies
\begin{equation}\label{twistedcomm}
\begin{aligned}
\{Q,Q^{12}\}&\propto P_{\zb}\\
\{Q,Q_{\bt}^{~22}\}&\propto P_{\beta},\\
\end{aligned}
\end{equation}
 where $Q=Q_{22}$. This indicates a topological-holomorphic theory, i.e., one where correlation functions of $Q$-invariant local operators have holomorphic dependence on $\Sigma$ and no other dependence on $M$.
 
 Next, setting the parameter $\zeta_{11}$ equal to 1, 
the 
corresponding supersymmetry transformations are
\begin{equation}\label{twistedtrans}
\begin{aligned}
Q\cA_{\al}&=0\\
Q\cAb_{\al}&=-8\psi_{\al 22}\\
QA_{\zb}&=0\\
QA_{z}&=4\eta_{12}\\
Q\phi_{\zhb}&=0\\
Q\phi_{\zh}&=-4i\eta_{21}\\
Q\eta_{11}&=2iF_{z\ov{z}}+2i[\phi_{\wht{z}},\phi_{\ov{\wht{z}}}]-D_{\beta}\phi^{\beta}\\
Q\eta_{12}&=0\\
Q\eta_{21}&=0\\
Q\eta_{22}&=4D_{\zb}\phi_{\zhb}\\
Q\psi_{\al 11}&=-\frac{1}{2}\varepsilon^{\bt\gm}_{~~~\al}{\cF}_{\bt\gm}\\
Q\psi_{\al 12}&=2\cD_{\al}\phi_{\zhb}\\
Q\psi_{\al 21}&=-i2{\cF}_{\al \zb}\\
Q\psi_{\al 22}&=0,\\
\end{aligned}
\end{equation}
where we have defined the complex coordinates $z=x^4+ix^5$ and $\ov{z}=x^4-ix^5$, the complex gauge fields
\begin{equation}
\begin{aligned}
\cA_{\al}=A_{\al}+i\phi_{\al}, && \cAb_{\al}=A_{\al}-i\phi_{\al},
\end{aligned}
\end{equation}
and
\begin{equation}
\begin{aligned}
A_{z}=\frac{1}{2}(A_{4}-iA_5), && A_{\zb}=\frac{1}{2}(A_{4}+iA_5),
\end{aligned}
\end{equation}
whereby we have the covariant derivatives
\begin{equation}
\begin{aligned}
\cD_{\al}=\del_{\al}+[\cA_{\al},\cdot\textrm{ }], && \cDb_{\al}=\del_{\al}+[\cAb_{\al},\cdot\textrm{ }],
\end{aligned}
\end{equation}
and 
\begin{equation}
\begin{aligned}
D_{z}=\del_{z}+[A_{z},\cdot\textrm{ }], && D_{\zb}=\del_{\zb}+[A_{\zb},\cdot\textrm{ }],
\end{aligned}
\end{equation}
and the field strengths $\cF_{\bt\gm}=[\cD_{\bt},\cD_{\gm}]$, $\cF_{\al z}=[\cD_{\al},D_{z}]$ and $F_{z \zb}=[D_{z},D_{\zb}]$.
We have also defined the scalar fields 
 \begin{equation}
\begin{aligned}
\phi_{\zh}=\frac{1}{2}(\phi_{\wht{4}}-i\phi_{\wht{5}}), && \phi_{\zhb}=\frac{1}{2}(\phi_{\wht{4}}+i\phi_{\wht{5}}).
\end{aligned}
\end{equation}
Note that these transformations leave invariant the NS5 boundary conditions, which take the following form after twisting: 
\begin{equation}\label{twistedbc}
\begin{aligned}
A_3=0|_{x^3=0}, && \del_3A_{\mu}=0|_{x^3=0},\\
\del_3\phi^{1,2}=0|_{x^3=0}, && \phi^{{3},\wht{4},\wht{5}}=0|_{x^3=0},\\
\del_3\eta_{1\mbh}=0|_{x^3=0}, && \eta_{2\mbh}=0\bc,\\
\psi_{\til{\al}1\mbh}=0\bc,&&\del_3\psi_{\til{\al}2\mbh}=0\bc, \\ \del_3\psi_{31\mbh}=0\bc,&&\psi_{32\mbh}=0\bc,
\end{aligned}
\end{equation}
where $\til{\al}=1,2$.

Now, let us proceed to reexpress the action in a form suitable for localization. The partially twisted action we obtain can be put in the form
\begin{equation}\label{twisted}
\begin{aligned}
S_{twisted}= -\frac{1}{g_5^2} \int_M d^5x ~\Tr\Big(&\frac{1}{4}{\cF}^{\beta\gamma}\ov{\cF}_{\beta\gamma} + 2{\cF}^{\alpha}_{~\ov{z}}\ov{\cF}_{\alpha z} +2\cD^{\alpha}\phi_{\ov{\wht{z}}}\ov{\cD}_\alpha \phi_{\wht{z}}+8{D}_{{z}} \phi_{\wht{z}}D_{\zb}\phi_{\ov{\wht{z}}}
\\&+\frac{1}{2}(2iF_{z\ov{z}}+2i[\phi_{\wht{z}},\phi_{\ov{\wht{z}}}]-D_{\beta}\phi^{\beta})^2\\&
-4\veps^{\bt\gm\al}\psi_{\al 11}\cDb_{\bt}\psi_{\gm 22}+4\veps^{\alpha\beta}_{~~~\gamma}\psi^{\gamma}_{~21}\cD_{\al}\psi_{\bt 12}-i4\eta_{11}\cD_{\al}\psi^{\al}_{~22} \\&-i4\psi^{\al}_{~21}\cDb_{\al}\eta_{12}+i4\psi^{\al}_{~12}\cDb_{\al}\eta_{21}+i4\eta_{22}\cD_{\al}\psi^{\al}_{~11}\\&-i8\psi^{\gm}_{~21}D_z\psi_{\gm 22}-i8\psi^{\gm}_{~11}D_{\zb}\psi_{\gm 12}+i8\eta_{22}D_z\eta_{21}+i8\eta_{11}D_{\zb}\eta_{12}\\&
-8\psi^{\bt}_{~12}[\phi_{\zh},\psi_{\bt 22}]-8\psi^{\bt}_{~21}[\phi_{\zhb},\psi_{\bt 11}]+8\eta_{22}[\phi_{\zh},\eta_{12}]+8\eta_{11}[\phi_{\zhb},\eta_{21}]\Big).
\end{aligned}
\end{equation}
In obtaining the form of the action given in \eqref{twisted}, we have performed several integration-by-parts, where we have used the NS5 boundary conditions \eqref{twistedbc}. In particular, the terms with only bosonic fields are equivalent to partial twists of the standard terms, i.e.,
\begin{equation}\label{standard}
\begin{aligned}
S_{boson}=-\frac{1}{g_5^2} \int_M d^5x~\Tr\bigg(&\frac{1}{4}F_{\al\bt}F^{\al\bt}+\frac{1}{4}F_{\al n}F^{\al n}+\frac{1}{4}F_{m\bt}F^{m\bt}+\frac{1}{4}F_{mn}F^{mn}\\& +\frac{1}{2}D^{\al}\phi^{\bt}D_{\al}\phi_{\bt}+\frac{1}{2}D^{\al}\phi^{\hn}D_{\al}\phi_{\hn}\\&+\frac{1}{2}D^m\phi^{\bt}D_m\phi_{\bt}+\frac{1}{2}D^m\phi^{\hn}D_m\phi_{\hn}\\&+\frac{1}{4}[\phi^{\al},\phi^{\bt}][\phi_{\al},\phi_{\bt}]+\frac{1}{4}[\phi^{\al},\phi^{\hn}][\phi_{\al},\phi_{\hn}]\\&+\frac{1}{4}[\phi^{\hm},\phi^{\bt}][\phi_{\hm},\phi_{\bt}]+\frac{1}{4}[\phi^{\hm},\phi^{\hn}][\phi_{\hm},\phi_{\hn}]\bigg).
\end{aligned}
\end{equation}
  At this point, we note that the supersymmetry transformations \eqref{twistedtrans} satisfy $Q^2=0$ on-shell, that is, with the aid of the equation of motion obtained from varying $\eta_{11}$. In order for $Q^2=0$ to be satisfied off-shell, we ought to introduce an auxiliary field to the action, i.e.,
\begin{equation}\begin{aligned}
&-\frac{1}{g_5^2} \int_M d^5x ~\Tr \bigg(\frac{1}{2}\big(2iF_{z\ov{z}}+2i[\phi_{\wht{z}},\phi_{\ov{\wht{z}}}]-D_{\beta}\phi^{\beta}\big)^2 \bigg)\\ \rightarrow &-\frac{1}{g_5^2} \int_M d^5x ~\Tr \bigg(d\big(2iF_{z\ov{z}}+2i[\phi_{\wht{z}},\phi_{\ov{\wht{z}}}]-D_{\beta}\phi^{\beta}\big) -\frac{1}{2}d^2\bigg),
\end{aligned}
\end{equation}
and modify \eqref{twistedtrans} such that 
\begin{equation}
\begin{aligned}
Q\eta_{11}&=d\\
Qd&=0.
\end{aligned}
\end{equation}

Then, the action \eqref{twisted} can be written in terms of a $Q$-exact part and a $Q$-invariant part, i.e.,
\begin{equation}\label{qexact}
\begin{aligned}
S_{twisted}=Q\Psi -\frac{1}{g_5^2} \int_M d^5x ~\Tr\bigg(& 4 \veps^{\al\bt\gm}\psi_{{\gm}21}\cD_{\al}\psi_{\bt 12}+i2\veps^{\al\rho\si}\eta_{22}\cD_{\al}\chi_{\rho\si 11}-i4\veps^{{\gm}\rho\si}\chi_{\rho\si 11}D_{\zb}\psi_{{\gm}12}\\&-4\veps^{\bt \rho\si}\psi_{\bt 21}[\phi_{\zhb},\chi_{\rho\si 11}]\bigg),
\end{aligned}
\end{equation}
where
\begin{equation}
\begin{aligned}
\Psi=-\frac{1}{g_5^2} \int_M d^5x ~\Tr\bigg(&-\frac{1}{4}\chi^{\bt\gm}_{~~11}\cFb_{\bt\gm}+i\psi^{\al}_{~21}\cFb_{\al z}+\psi^{\al}_{~12}\cDb_{\al}\phi_{\zh}+2\eta_{22}D_z \phi_{\zh}\\&+\eta_{11}\big(2iF_{z\ov{z}}+2i[\phi_{\wht{z}},\phi_{\ov{\wht{z}}}]-D_{\beta}\phi^{\beta}\big)-\frac{1}{2}\eta_{11}d\bigg).
\end{aligned}
\end{equation}
Here, we have performed the field redefinition
 \begin{equation}
\chi_{\rho\si 11}=\veps_{\rho \si \al}\psi^{\al}_{~11},
\end{equation}
where $\chi_{\rho\si 11}$ are the components of a 2-form on $V$, satisfying $Q\chi_{\rho\si 11}= - \cF_{\rho\si}$.  The $Q$-invariance of the $Q$-exact term in \eqref{qexact} follows since $Q$ is nilpotent, while the remaining terms are $Q$-invariant due to the Bianchi identities 
\begin{equation}
\cD_{\al}\cF_{\bt\gm}+\cD_{\bt}\cF_{\gm\al}+\cD_{\gm}\cF_{\al\bt}=0
\end{equation}
and 
\begin{equation}
D_{\zb}\cF_{\bt\gm}+\cD_{\bt}\cF_{\gm \zb}+\cD_{\gm}\cF_{\zb\bt}=0.
\end{equation}

Note that the dependence on the metric of $V$ is completely contained in the $Q$-exact term in \eqref{qexact}, while there is dependence on the complex structure of $\Sigma$ 
which can be observed via the presence of the derivative $D_{\zb}$ in one of the non-$Q$-exact terms.
Hence, along the boundary, $\del M=Y\times \Sigma$, the partially twisted theory we have derived is topological along $Y$ but depends on the complex structure of $\Sigma$, just as in Costello's 4d Chern-Simons theory.
\subsection{Boundary action}
In general, the low energy worldvolume action of coincident D4-branes admits a topological term which couples to the RR 1-form, $R$, sourced by D0-branes in the worldvolume, i.e., 
\begin{equation}\label{topo}
S_{top}=\frac{i}{g_5^2}\int_M R  \wedge\Tr ( F \wedge F).
\end{equation}
In the following, we shall only be concerned with D0-branes which are charged such that the RR 1-form $R$ is closed, which allows us to write \eqref{topo} as a boundary action.

In order to include such a coupling without breaking the supersymmetry generated by $Q$, we shall first require that  $R$ has meromorphic dependence on $\Sigma$, i.e., $R=C(z)dz$. Upon doing so, the boundary RR 1-form coupling can be written as
\begin{equation}\label{CSC}
\frac{-i}{g_5^2}\int_{\del M} C\wedge \textrm{Tr}\bigg(A\wedge dA + \frac{2}{3} A\wedge A\wedge A\bigg),
\end{equation}
where $C=C(z)dz$.
Now, this boundary action only depends on $A_1$, $A_2$ and $A_{\zb}$, where 
$A_{\zb}$ is $Q$-invariant. Hence, in order to preserve $Q$-invariance along the boundary, we 
add boundary interaction terms involving $\phi_1$ and $\phi_2$ to the action, such that the dependence on ($A_1$,$A_2$) is replaced precisely by dependence on ($\cA_1$,$\cA_2$), resulting in the $Q$-invariant boundary action 
\begin{equation}\label{CSCC}
S_{boundary}=\frac{-i}{g_5^2}\int_{\del M} C\wedge \textrm{Tr}\bigg(\mA\wedge d\mA + \frac{2}{3} \mA\wedge \mA\wedge \mA\bigg),
\end{equation}
where we have used the notation $A_{\zb}=\cA_{\zb}$.

\subsection{Localization}
We shall now specialize to the case where $Y$ is a framed 2-manifold, the most important examples (in the context of the Yang-Baxter equation) that we will consider being $T^2$ and $\R^2$, and where $\Sigma$ is a Riemann surface which is either $\C$, $\C^{\times}$ or $\C/(\Z+\tau\Z)$. 
These three choices of Riemann surfaces will eventually correspond to rational, trigonometric and elliptic integrable lattice models. 

To evaluate the path integral, we shall rescale the $Q$-exact part of the action by a parameter $s$, which we shall eventually take to be very large. Since the path integral localizes to the fixed points of the fermionic symmetry, it will
be convenient to evaluate it by expanding in perturbation theory around these fixed points. 

Let us first integrate out the auxiliary field $d$. Then, we gauge $A_3$ away to zero. 
Next, denoting solutions of the bosonic fixed points of the fermionic symmetry as $X_0$ and fluctuations around these points as $\widetilde{X}$ for any bosonic field $X$, we expand these fields as $X_0+\til{X}$, and we rescale the fluctuations and the fermionic fields as follows:
\begin{equation}
\begin{aligned}
\til{\cA}_{\til{\al}}, \ov{\til{\cA}}_{\til{\al}}, \til{A}_z, \til{A}_{\zb}, \til{\phi}_{\zh}, \til{\phi}_{\zhb}, \til{\phi}_3 &\rightarrow \frac{\til{\cA}_{\til{\al}}}{s}, \frac{\ov{\til{\cA}}_{\til{\al}}}{s}, \frac{\til{A}_z}{s^2}, \frac{\til{A}_{\zb}}{s^2}, \frac{\til{\phi}_{\zh}}{s^2}, \frac{\til{\phi}_{\zhb}}{s^2}, \frac{\til{\phi}_3}{s^{\frac{7}{2}}}\\
\chi_{3\tal 11}, \psi_{\tal 22}, \psi_{312}, \psi_{321}, \psi_{322}, \eta_{11}, \eta_{12}, \eta_{21} &\rightarrow \frac{\chi_{3\tal 11}}{s}, \frac{\psi_{\tal 22}}{s}, \frac{\psi_{312}}{s^2}, \frac{\psi_{321}}{s^{5/2}}, \frac{\psi_{322}}{s^{5/2}}, \frac{\eta_{11}}{s^{3/2}}, \frac{\eta_{12}}{s^{3/2}}, \frac{\eta_{21}}{s^{2}}
\end{aligned}
\end{equation}
(here, and in what follows, ($\tal,\tbt,\tgm,\dots)=1,2$).
Since the theory is topological along $V=Y\times \R_+$, we can also rescale the inverse of the metric as
\begin{equation}
\begin{aligned}
g^{33}, g^{\til{\al}\til{\bt}} \rightarrow s^{3}g^{33},  \frac{1}{s^2}g^{\til{\al}\til{\bt}},
\end{aligned}
\end{equation}
where $g_{\til{\al}\til{\bt}}$ is the flat metric along $Y$.

Taking $s\rightarrow \infty$, the total action becomes 
\begin{equation}
\begin{aligned}
S=-\frac{1}{g_5^2} \int_M d^5x ~\Tr \bigg(& \frac{1}{2}g^{33}g^{\tgm\tbt}\del_{3}\til{\cA}_{\tbt}\del_{3}\ov{\til{\cA}}_{\tgm}+2g^{33}\del_{3}\til{A}_{\zb}\del_{3}\til{A}_z\\&+2g^{33}\del_{3}\til{\phi}_{\zhb}\del_{3}\til{\phi}_{\zh}+\frac{1}{2}g^{33}g^{33}\del_3\til{\phi}_3\del_3\til{\phi}_3\\& -4g^{33}g^{\tgm\tbt}\chi_{3\tbt 11}\del_{3}\psi_{\tgm 22}-i4g^{33} \eta_{11}\del_{3}\psi_{322}\\&-i4g^{33}\psi_{321}\del_{3}\eta_{12}+i4g^{33}\psi_{312}\del_{3}\eta_{21}\\& +4\veps^{3\tbt\tgm}\psi_{\tgm 21} \del_{3}\psi_{\tbt 12}+i2 \veps^{3 \tbt \tgm}\eta_{22}\del_{3}\chi_{\tbt\tgm 11}\bigg) \\-\frac{i}{g_5^2}\int_{\del M} C\wedge \textrm{Tr}&\bigg(\mA_0\wedge d\mA_0 + \frac{2}{3} \mA_0\wedge \mA_0\wedge \mA_0\bigg)
\end{aligned}
\end{equation}
where 
we have used the fact that $\phi_{30}$, $\phi_{\zh 0}$ and $\phi_{\zhb 0}$ are equal to zero.\footnote{This follows from the equivalence between the terms with only bosonic fields in \eqref{twisted} and \eqref{standard}. The latter is a sum of squares, which implies that $\phi_{30}$, $\phi_{\zh 0}$ and $\phi_{\zhb 0}$ are each covariantly constant, and commute with all the other $\phi$ fields. The Dirichlet boundary conditions on these three fields then imply that they must be zero everywhere on $M$.}  
 Performing the path integral over the fluctuations and fermions,\footnote{We have assumed here that there are no fermionic zero modes. This is a reasonable assumption, given that the NS5 boundary conditions consist of Dirichlet and Neumann boundary conditions which are both elliptic in general, and the index of the fermionic operators ought to be proportional to $\chi(M)\textrm{dim }G$, which vanishes for our present choice of $M$.}
we obtain a constant factor which can be absorbed into the measure, giving us
\begin{equation}
\int \cD\cA_{\tal 0}{\cD}\cAb_{\tal 0}\cD A_{z 0}\cD A_{\zb 0}~ e^{\frac{i}{g_5^2}\int_{\del M} C\wedge\textrm{Tr}\big( \mA_0\wedge d\mA_0 + \frac{2}{3} \mA_0\wedge \mA_0\wedge \mA_0\big)}.
\end{equation}
Then, performing the path integral over $\cAb_{\tal 0}$ and renormalizing, we end up with 
\begin{equation}\label{box1}
\boxed{
\int_{\Gamma} \cD\cA_{\tal 0}\cD \cA_{z 0}\cD \cA_{\zb 0}~ e^{\frac{i}{g_5^2}\int_{\del M} C\wedge\textrm{Tr}\big( \mA_0\wedge d\mA_0 + \frac{2}{3} \mA_0\wedge \mA_0\wedge \mA_0\big)}}
\end{equation}
where we now use the notation $A_{z 0}=\cA_{z 0}$ and $A_{\zb 0}=\cA_{\zb 0}$ in the path integral measure.
This is 
the partition function of Costello's 4d Chern-Simons theory with gauge group $U(N)_{\C}=GL(N,\C)$ (where the Planck constant $\hbar=g_5^2$), with the functional integral performed over a nonperturbative integration cycle $\Gamma$ in field space, corresponding to the restriction to $\del M$ of the bosonic fixed points of the fermionic symmetry generated by $Q$, i.e.,
\begin{equation}\label{box2}
\boxed{
\begin{aligned}
\cF_{\tal \tbt 0}&=0  \\
\cF_{\tal \zb 0}&=0\\
2i\cF_{z\zb 0}-D_{\tal 0}\phi_0^{\tal}&=0
\end{aligned}}
\end{equation}
As shown recently in \cite{CostelloYagi}, one can also obtain 
4d Chern-Simons theory from string theory by realizing it via a stack of D5-branes supported on the product of an $\Omega$-background disk and $Y\times \Sigma$. 
\subsection{Wilson lines}
Now, it is a fact that $Q\cA_{\tal}$=0 also allows us to define supersymmetric Wilson lines along $Y$, i.e.,
\begin{equation}
W=\textrm{Tr}(P~e^{\int_{L \subset Y}\cA}),
\end{equation}
as observables of the 5d topological-holomorphic theory that are associated with representations of the complex Lie algebra $\mathfrak{g}=\mathfrak{gl}(N,\C)$. From the point of view of string theory, these Wilson lines
arise from the worldsheet boundaries of fundamental strings ending on the D4-brane worldvolume. 

Now, in \cite{CWY}, a general class of Wilson lines was considered,
  i.e., not only those associated with representations of $\mathfrak{g}$, but also representations of $\mathfrak{g}[[z]]$ (the Lie algebra of polynomial loops of $\mathfrak{g}$). These Wilson lines are necessary in deriving the Yangian algebra associated with rational integrable lattice models. Such a Wilson line associated with $\mathfrak{g}[[z]]$ is constructed by starting with a Wilson line along $Y$ in a representation of $\mathfrak{g}$, and giving holomorphic dependence on $\Sigma$ to the gauge field in the operator, while also removing the trace (gauge invariance is maintained by taking $Y$ to be very large, 
 and insisting that the gauge field vanishes at infinity
). In the string picture, such a Wilson line ought to be realized by an identical modification to the Wilson line along the boundary of a fundamental string worldsheet, i.e., choosing a background gauge field with 
($z,\zb$)-dependence, and which vanishes at infinity along $\Sigma$. Note that since $P_{\zb}$ is $Q$-exact (see \eqref{twistedcomm}), the $\zb$-dependence is actually trivial in the sector of the worldvolume theory we are studying, and therefore the Wilson line realized along $Y$ has holomorphic $z$-dependence.


Let us consider 
the path integral of the 5d topological-holomorphic theory with Wilson lines along $Y\subset\del M$. Localization of the path integral along the lines of the previous subsection leads us to Costello's 4d Chern-Simons theory with 
 Wilson line insertions in the path integral, i.e.,
\begin{equation}\label{superwil}
\boxed{
\int_{\Gamma}  \cD\cA_{\tal 0}\cD \cA_{z 0}\cD \cA_{\zb 0} \prod_i\textrm{Tr}(P~e^{\int_{L_i}\cA_0}) ~e^{\frac{i}{g_5^2}\int_{\del M} C\wedge\textrm{Tr}\big( \cA_0\wedge d\cA_0 + \frac{2}{3} \cA_0\wedge \cA_0\wedge \cA_0\big)}}
\end{equation} 
  and for Wilson lines associated with representations of $\mathfrak{g}=\mathfrak{gl}(N,\C)$
 forming a lattice along $Y\subset\del M$, this correlation function is identified with the partition function of an integrable lattice model. This is because contracting R-matrices (which correspond to 
  intersections of Wilson lines) and taking a trace gives a transfer matrix, and similarly contracting transfer matrices and taking a trace gives us the partition function.

\section{Categorification of R-matrix elements}

\subsection{T-duality as a lift to 6d}

To categorify the R-matrix elements and the Yang-Baxter equation with spectral parameter, we need to lift our 5d partially topological theory to a 6d one. From the string theory perspective, this corresponds to T-duality along one of the directions transverse to the D4-brane worldvolume. For our purposes, we shall require that this direction lie along the NS5-brane, so that we end up with a D5-NS5 brane system. We shall pick this direction to be $x^6$, leading to the following type IIB brane configuration:
\begin{center}
\begin{tabular}{@{}lllllllllll@{}}
\toprule
   & \textbf{1} & \textbf{2} & \textbf{3} & \textbf{4} & \textbf{5} & \textbf{6} & \textbf{7} & \textbf{8} & \textbf{9} & \textbf{10}\\
   \textbf{D5} & \x & \x & \x & \x & \x & \x & & & &\\
   \textbf{NS5} & \x & \x &  & \x & \x & \x & \x & & & \\
\bottomrule
 \end{tabular}
\end{center}
However, recall that the partial twisting in Section 2.1 was induced by taking the  $\{x^6,x^7,x^8\}$ directions to form the normal bundle to $\til{V}\subset T^*\til{V}$, where $\til{V}=Y\times \R$ lies along the $\{x^1,x^2,x^3\}$ directions. Hence, in order to facilitate the T-duality along the $x^6$ direction, we ought to replace the $\R$ fiber in this direction by a very large circle. T-dualizing the D4-NS5 system then leads to the above D5-NS5 system, where $x^6$ is a local coordinate of a very small circle.

Let us now describe this T-duality as a 6d lift of the 5d action. Firstly, we note that the lift along the $x^6$ direction implies that the field $\phi_1$ is replaced by the gauge field component $A_6$. This in turn implies that 
\begin{equation}\label{lift1}
\begin{aligned}
\cA_1=A_1+i\phi_1 &\rightarrow A_1+iA_6 = 2A_{\wb}\\
\cAb_1=A_1-i\phi_1 &\rightarrow A_1-iA_6 = 2A_{w},
\end{aligned}
\end{equation}
where we have defined the complex coordinates $w=x^1+ix^6$ and $\wb=x^1-ix^6$. 

We shall first lift the boundary action \eqref{CSCC} using the first identification in \eqref{lift1}, as well as $\del_1\rightarrow 2\del_{\wb}$.\footnote{This choice, as opposed to $\del_1\rightarrow 2\del_{w}$, is necessary to obtain a boundary action invariant under (small) gauge transformations.}  Upon doing so, we obtain
\begin{equation}\label{CSCClifted}
\begin{aligned}
&\frac{-i}{g_5^2}\int_{\del M} C\wedge \textrm{Tr}\bigg(\mA\wedge d\mA + \frac{2}{3} \mA\wedge \mA\wedge \mA\bigg)\\
\rightarrow & \frac{-1}{g_5^2 (2\pi r)}\int_{\del M \times S^1} dw\wedge C\wedge \textrm{Tr}\bigg(\mA\wedge d\mA + \frac{2}{3} \mA\wedge \mA\wedge \mA\bigg)
\end{aligned}
\end{equation}
where the circle with respect to which we are lifting has radius $r$, and where we use the notation $A_{\wb}=\cA_{\wb}$. 

Next, we wish to lift the bulk action. To do this, we shall first lift the $Q$-transformations of the 5d action, via 
\begin{equation}\label{lift2}
\begin{aligned}
\cD_1 \rightarrow 2D_{\wb}\\
\cDb_1 \rightarrow  2D_{w}.
\end{aligned}
\end{equation}
The $Q$-transformations we obtain in this way are
\begin{equation}\label{liftedtrans}
\begin{aligned}
Q\cA_{i}&=0\\
Q\cAb_{i}&=-8\psi_{i 22}\\
QA_{\wb}&=0\\
QA_{w}&=-4\teta_{22}\\
QA_{\zb}&=0\\
QA_{z}&=4\eta_{12}\\
Q\phi_{\zhb}&=0\\
Q\phi_{\zh}&=-4i\eta_{21}\\
Q\eta_{11}&=d\\
Qd&=0\\
Q\eta_{12}&=0\\
Q\eta_{21}&=0\\
Q\eta_{22}&=4D_{\zb}\phi_{\zhb}\\
Q\tpsi_{j 11}&= 2\cF_{j \wb}\\
Q\chi_{ij 11}&= - \cF_{ij}\\
Q\teta_{ 12}&=4D_{\wb}\phi_{\zhb}\\
Q\psi_{i 12}&=2\cD_{i}\phi_{\zhb}\\
Q\teta_{ 21}&=-i4{F}_{\wb \zb}\\
Q\psi_{i 21}&=-i2{\cF}_{i \zb}\\
Q\teta_{ 22}&=0,\\
Q\psi_{i 22}&=0,\\
\end{aligned}
\end{equation}
where $i,j,k,\ldots=2,3$, and where we have defined the fields $\tpsi_{j11}=\chi_{1j11}$, $\teta_{12}=\psi_{112}$, $\teta_{21}=\psi_{121}$, and $\teta_{22}=\psi_{122}$.
Note that the boundary action \eqref{CSCClifted} remains $Q$-invariant after lifting.
We can now directly lift the bulk action in the form \eqref{qexact} to 6d, which gives 
\begin{equation}\label{qexact6d}
\begin{aligned}
S^{6d}_{twisted}=Q\Psi' -\frac{1}{g_5^2 2\pi r} \int_{M\times S^1} d^6x ~\Tr\bigg(& 8 \veps^{jk}\psi_{{k}21}D_{\wb}\psi_{j 12}-4 \veps^{ik}\psi_{{k}21}\cD_{i}\teta_{12}+4 \veps^{ij}\teta_{21}\cD_{i}\psi_{j 12}\\&+i4\veps^{jk}\eta_{22}D_{\wb}\chi_{jk 11}-i4\veps^{ij}\eta_{22}\cD_{i}\tpsi_{j 11}\\&-i4\veps^{jk}\chi_{jk 11}D_{\zb}\teta_{12}+i8\veps^{ij}\tpsi_{j 11}D_{\zb}\psi_{i 12}\\&-4\veps^{jk}\teta_{ 21}[\phi_{\zhb},\chi_{jk 11}]+8\veps^{ij}\psi_{i 21}[\phi_{\zhb},\tpsi_{j 11}]\bigg),
\end{aligned}
\end{equation}
where $\veps^{1jk}=\veps^{jk}$, and where
\begin{equation}
\begin{aligned}
\Psi'=-\frac{1}{g_5^2 2\pi r} \int_{M\times S^1} d^6x ~\Tr\bigg(&-\frac{1}{4}\chi^{ij}_{~~11}\cFb_{ij}+\tpsi^{j}_{11}\cFb_{j w}+i\psi^{i}_{~21}\cFb_{i z}+i2\teta_{21}F_{wz}\\&+\psi^{i}_{~12}\cDb_{i}\phi_{\zh}+2\teta_{12}D_w \phi_{\zh}+2\eta_{22}D_z \phi_{\zh}\\&+\eta_{11}\big(2iF_{z\ov{z}}+2iF_{w\ov{w}}+2i[\phi_{\wht{z}},\phi_{\ov{\wht{z}}}]-D_{j}\phi^{j}\big)-\frac{1}{2}\eta_{11}d\bigg).
\end{aligned}
\end{equation}
We may verify that this is (a partially twisted version of) 6d $\mathcal{N}=(1,1)$ Super Yang-Mills by 
studying the terms with only bosonic fields in \eqref{qexact6d}, i.e.,
\begin{equation}\label{6db1}
\begin{aligned}
S^{6d}_{boson}=-\frac{1}{g_5^2 2\pi r} \int_{M\times S^1} d^6x ~\Tr\bigg(&\frac{1}{4}\cF^{ij}\cFb_{ij}+2\cF^{i}_{~\zb}\cFb_{iz}+2\cF^{i}_{~\wb}\cFb_{iw}+8F_{\wb\zb}F_{wz}\\&+2\cD^{i}\phi_{\zhb}\cDb_i\phi_{\zh}+8D_{\zb}\phi_{\zhb}D_z\phi_{\zh}+8D_{\wb}\phi_{\zhb}D_w\phi_{\zh}\\&+d\big(2iF_{z\ov{z}}+2iF_{w\ov{w}}+2i[\phi_{\wht{z}},\phi_{\ov{\wht{z}}}]-D_{j}\phi^{j}\big)-\frac{1}{2}d^2\bigg).
\end{aligned}
\end{equation}
Integrating $d$ out of the action, this can be reexpressed as 
\begin{equation}\label{6db2}
\begin{aligned}
S^{6d}_{boson}=-\frac{1}{g_5^2 2\pi r} \int_{M\times S^1} d^6x ~\Tr\bigg(&\frac{1}{4}F_{xy}F^{xy}+\frac{1}{4}F_{iy}F^{iy}+\frac{1}{4}F_{xj}F^{xj}+\frac{1}{4}F_{ij}F^{ij}\\& +\frac{1}{2}D^i\phi^jD_i\phi_j+\frac{1}{2}D^i\phi^{\hn}D_i\phi_{\hn}\\&+\frac{1}{2}D^x\phi^jD_x\phi_j+\frac{1}{2}D^x\phi^{\hn}D_x\phi_{\hn}\\&+\frac{1}{4}[\phi^i,\phi^j][\phi_i,\phi_j]+\frac{1}{4}[\phi^i,\phi^{\hn}][\phi_i,\phi_{\hn}]\\&+\frac{1}{4}[\phi^{\hm},\phi^j][\phi_{\hm},\phi_j]+\frac{1}{4}[\phi^{\hm},\phi^{\hn}][\phi_{\hm},\phi_{\hn}]\bigg)
\end{aligned}
\end{equation}
after integration by parts (making use of the NS5 boundary condition $\phi_3=0$), where $x,y=z,\zb,w,\wb$. Hence, the 6d theory \eqref{qexact6d}, together with the boundary action
\begin{equation}
\label{6dboundaryaction}
S^{6d}_{boundary}=\frac{-1}{g_5^2 (2\pi r)}\int_{\del M \times S^1} dw\wedge C\wedge \textrm{Tr}\bigg(\mA\wedge d\mA + \frac{2}{3} \mA\wedge \mA\wedge \mA\bigg)
\end{equation}
 we obtained by lifting, is 6d maximally supersymmetric Yang-Mills theory partially twisted along the $x^2$ and $x^3$ directions, with a boundary coupling to the RR 2-form $idw\wedge C$. It can be observed from $\eqref{qexact6d}$ that the dependence on the metric in the $x^2$ and $x^3$ directions is completely contained in the $Q$-exact term, while some of the remaining terms depend on the complex structures of $\Sigma$ or $\Sigma'$, where $\Sigma'$ is the Riemann surface with complex coordinates $w$ and $\wb$.

As an aside, we note that using similar arguments to section 2.2, one ought to be able to localize this partially twisted theory to a 5d Chern-Simons theory
with an action of the form given in \eqref{6dboundaryaction}, with the path integral performed over the nonperturbative integration cycle $\Gamma'$ defined by 
\begin{equation}
\begin{aligned}
\cF_{2\zb 0}&=0\\
\cF_{2\wb 0}&=0\\
\cF_{\wb\zb 0}&=0\\
2i\cF_{z\zb 0}+2i\cF_{w\wb 0}-D_{20}\phi_0^{2}&=0,
\end{aligned}
\end{equation}
which happen to be the lifts of the equations given in \eqref{box2}.
For $C=dz$, this is just the commutative limit of Costello's noncommutative 5d Chern-Simons theory studied in \cite{Costello2,Costello3,ACMV}, which was obtained from a stack of D6-branes supported on the product of an $\Omega$-background plane and $\R\times \C^2$. Indeed, deforming the $\Omega$-background plane to an infinitely long cigar and applying T-duality along its circle fibers, we obtain the D5-NS5 system. The reason we obtain a commutative version is that noncommutativity only arises in the presence of a particular nonzero $B$-field \cite{Costello2}, which we do not have. 


\subsection{S-duality}

We now go one step further, and study the S-dual of this system, whereby D5 and NS5-branes are exchanged. We note that both D5-branes and NS5-branes in type IIB string theory are described at low energy by 6d $\mathcal{N}=(1,1)$ Super Yang-Mills theory. Hence, the dual theory is also a twist of 6d $\mathcal{N}=(1,1)$ Super Yang-Mills, but one which lives on a stack of NS5-branes ending on a D5-brane.
In addition, the F-strings ending on the D5-branes, which gave rise to Wilson lines, become D1-branes ending on the NS5-branes, which are also described by Wilson lines. Also, the background metric is rescaled under S-duality by a factor of $1/g$, where $g$ is the type IIB string coupling, and therefore variables with dimensions of length are rescaled by $\sqrt{g}$. 

Next, we note that before S-duality, the coupling of the
5d boundary action of the D5-brane worldvolume theory is $g_5^2(2\pi r)=(2\pi)^3g \alpha'$, 
and S-duality results in the
5d boundary action of the NS5-brane worldvolume theory having the coupling $(2\pi)^3\alpha'/g$. In other words, the boundary action of the S-dual theory is multiplied by $g/(2\pi)^3\alpha'=g_5^2/(2\pi)^5(\alpha')^{2}r^{-1}$. In addition, the coupling of the 6d bulk action of the  D5-brane worldvolume theory, which is $(2\pi)^3g \alpha'$ before S-duality, is replaced by the coupling $(2\pi)^3 \alpha'$ of the bulk action of the NS5-brane worldvolume after S-duality. Finally, the RR 2-form $idw\wedge C$ in the D5-brane worldvolume is identified with an NS-NS 2-form in the NS5-brane worldvolume theory.

We shall use this S-dual 6d theory to categorify the R-matrix elements, thereby providing a categorification of the Yang-Baxter equation with spectral parameter. In other words, we want to be able to describe the R-matrix elements in terms of the Hilbert space of a 6d theory. The reason for using the S-dual theory is that it allows us to write its path integral as a trace (over its supersymmetric Hilbert space) of an expression with an obvious expansion in positive powers of  $\hbar$, which the R-matrix is known to admit in the semi-classical limit. 




\subsection{Hilbert space of 6d theory and categorification}
The BPS equations to which the NS5-brane worldvolume theory localizes 
will be identical to those 
of the original 6d theory on the D5-branes, which can be obtained from \eqref{liftedtrans}, i.e.,
\begin{equation}\label{6dbps}
\boxed{
\begin{aligned}
\cF_{ij}&=0\\
{\cF}_{i \zb}&=0\\
\cF_{i \wb}&=0\\
{F}_{\wb \zb}&=0\\
2iF_{z\ov{z}}+2iF_{w\ov{w}}-D_{j}\phi^{j}&=0
\end{aligned}
}
\end{equation}
where we have integrated out $d$, and used the fact that equivalence between \eqref{6db1} and \eqref{6db2} and the Dirichlet boundary conditions on $\phi_{\zhb}$ imply that its BPS configurations are forced to be zero.

Let us now use these equations to describe the supersymmetric Hilbert space of the S-dual theory, taking the sixth dimension $S^1$ to be the Euclidean time dimension. We shall first find the space of classical ground states, which in the present case is a time-independent classical solution of the six-dimensional equations given in \eqref{6dbps}. That is, a classical ground state would solve the 5d  equations 
\begin{equation}\label{5dbps}
\boxed{
\begin{aligned}
{\cF}_{\bt\gm}&=0\\
{\cF}_{\al \zb}&=0\\
2iF_{z\ov{z}}-D_{\beta}\phi^{\beta}&=0
\end{aligned}}
\end{equation}
which are the 
BPS equations from our study of the D4-brane worldvolume theory. We shall assume for simplicity in what follows that these equations have a finite set of solutions (modulo gauge transformations), and that they are all nondegenerate (i.e., when expanding around a particular solution, there are no bosonic zero modes). The latter condition implies that expansion around such a solution gives a single quantum state of vanishing energy, at least in perturbation theory.

We shall now take into account quantum corrections to the classical spectrum. Firstly, it is known that nonzero energy eigenstates of a supersymmetric Hamiltonian always occur in pairs. In perturbation theory, the supersymmetric spectrum is unaffected by quantum corrections, because we always expand around a single approximate ground state, which is obtained by quantizing the corresponding classical solution, and hence because in perturbation theory only this single state is accessible, it is impossible for it to pair up with another state to leave the spectrum of supersymmetric ground states.  

However, nonperturbatively, quantum tunnelling via `instantons' between such approximate ground states can lift a pair of ground states to a pair of excited states. The `instantons' in our case are solutions to the 6d equations \eqref{6dbps}, and interpolate the 5d solutions of \eqref{5dbps}. An approximate ground state $\psi_{\mathcal{I}}$, i.e., which perturbatively obeys $Q\psi_{\mathcal{I}}=0$, would now obey
\begin{equation}
Q\psi_{\mathcal{I}}=\sum_{\mathcal{J}\in S_5}m_{\mathcal{I}\mathcal{J}}\psi_{\mathcal{J}},
\end{equation}
(where $S_5$ is the space of solutions of \eqref{5dbps}) in the full quantum theory, where $m_{\mathcal{I}\mathcal{J}}$ is obtained by summing contributions from all `instantons' that interpolate between the solutions labelled by $\mathcal{I}$ and $\mathcal{J}$.
 Hence, the quantum Hilbert space of ground states, $\mathcal{H}$, of the 6d theory will be given by the cohomology of the operator $Q$, which is the Floer cohomology corresponding to the equations \eqref{6dbps}.
 

Now, when $Y=T^2$, the trace in the Wilson loop operators in \eqref{superwil} can be disregarded due to the diffeomorphism invariance along $Y$ in our original D4-NS5 setup, which allows us to take $Y$ to be very large (gauge invariance will follow if we insist that the gauge field vanishes at infinity). When $Y=\R^2$, removing the trace in this manner can be done without any rescaling of $Y$. In what follows, we shall also view the system under consideration at a fixed length scale, which allows us to set $\alpha'$ and $r$ to convenient constants.
Let us now consider \eqref{superwil} with two perpendicular Wilson lines along $Y$, which gives rise to the R-matrix. Since the path integral of the 6d theory obtained via the aforementioned string dualities is equal to an R-matrix element, we can write
\begin{equation}\label{boxR}
\boxed{
R^{12}_{IK,JL}(z_1,z_2)=\textrm{Tr}_{\mathcal{H}}\big((-1)^{\textrm{F}}e^{-{\hbar}P}W^1_{IJ}(z_1)W^2_{KL}(z_2)\big)}
\end{equation}
where $z_1$ and $z_2$ are the positions of the Wilson lines (denoted as $W^1$ and $W^2$) on $\Sigma$ and correspond to the spectral parameters which label the R-matrix, $I,J,K,L$ are basis elements of the representations of the Wilson lines,
$\textrm{F}$ is the fermion number operator, and where the operator  
\begin{equation}
\boxed{
P=i\int_{\del M} C\wedge \textrm{Tr}\bigg(\mA\wedge d\mA + \frac{2}{3} \mA\wedge \mA\wedge \mA\bigg) }
\end{equation}
as obtained from the boundary topological term of the S-dual action at a point in time.

In other words, \textit{we have identified a vector space with each R-matrix element, whereby the vector space is the Floer cohomology of the set of 6d partial differential equations \eqref{6dbps}}. We may compute each R-matrix element from the solutions of these equations, which are associated with the Hilbert space of the 6d theory (we shall leave this for future work). This may be viewed as an indirect categorification of the Yang-Baxter equation, but we may also categorify directly. This involves three Wilson lines labelled by spectral parameters $z_1$, $z_2$ and $z_3$ in the original D4-NS5 setup in Section 2, whereby the topological invariance along $Y$ allows us to reproduce the diagrammatic form of the Yang-Baxter equation by moving one of the Wilson lines. We then apply T- and S-duality to both sides of the equation, whereby, 
with the moved Wilson line denoted as $\til{W}$, we may represent the Yang-Baxter equation,
\begin{equation}
\sum_{O,P,Q}R^{12}_{NM,QO}(z_1,z_2)R^{13}_{QL,IP}(z_1,z_3)R^{23}_{OP,JK}(z_2,z_3)=\sum_{R,S,T}R^{23}_{ML,RT}(z_2,z_3)R^{13}_{NT,SK}(z_1,z_3)R^{12}_{SR,IJ}(z_1,z_2)
\end{equation}
 as
\begin{equation}\label{boxYB}
\boxed{\textrm{Tr}_{\mathcal{H}}\big((-1)^{\textrm{F}}e^{-{\hbar}P}W^1_{NI}(z_1)W^2_{MJ}(z_2)W^3_{LK}(z_3)\big)=\textrm{Tr}_{\mathcal{H}}\big((-1)^{\textrm{F}}e^{-{\hbar}P}W^1_{NI}(z_1)\til{W}^2_{MJ}(z_2)W^3_{LK}(z_3)\big)}
\end{equation} 

Moreover, we may, in a similar vein, categorify the Yangian algebra associated with rational integrable lattice models, in the form of the RTT relation. The latter is realized in the 4d Chern-Simons theory using three Wilson lines as well, but with one of them associated with a representation of $\mathfrak{g}[[z]]$ instead of $\mathfrak{g}$ \cite{CWY2}. Moving this Wilson line using the topological invariance along $Y$ gives rise to the RTT relation, and therefore an expression of the form \eqref{boxYB} categorifies the Yangian algebra as well.
It is expected that affine and elliptic quantum algebras can be categorified in an analogous manner.

\subsection{S-dual 4d Chern-Simons theory}
 

To obtain the 6d theory in Section 3.1, we applied T-duality along a very large circle in the $x^6$ direction, which resulted in a very small circle in the 6d worldvolume of the D5-branes. Hence, at low energies, the 6d theory (and its S-dual) can effectively be regarded as the 5d theory obtained via dimensional reduction.
Considering the S-dual 6d theory, we may localize the path integral of the effective 5d theory we obtain at low energies, in an analogous manner to how we localized the path integral of the 5d theory in Sections 2.3 and 2.4. Doing so, we obtain
\begin{equation}\label{lasteq}
\int_{\Gamma} \cD\cA_{\tal 0}\cD \cA_{z 0}\cD \cA_{\zb 0} \prod_i\textrm{Tr}(P~e^{\int_{L_i}\cA_0})~ e^{{i\hbar}\int_{ \del M} C\wedge\textrm{Tr}\big( \mA_0\wedge d\mA_0 + \frac{2}{3} \mA_0\wedge \mA_0\wedge \mA_0\big)},
\end{equation}
where $\Gamma$ is the cycle defined by the equations $\cF_{\tal \tbt 0}=0$, $\cF_{\tal \zb 0}=0$ and
$2i\cF_{z\zb 0}-D_{\tal 0}\phi_0^{\tal}=0$.
Hence, we have an `S-dual' of Costello's 4d Chern-Simons theory where the coupling is inverted as 
\begin{equation}
\frac{1}{\hbar}\rightarrow \hbar.
\end{equation}

This may be compared with the S-duality of analytically-continued 3d Chern-Simons theory, studied in \cite{TY,DimofteGukov}. S-duality of this Chern-Simons theory 
inverts the coupling and exchanges the gauge group with its Langlands dual, and was argued to arise as a consequence of the S-duality of 4d $\mathcal{N}=4$ SYM. 
Similarly, in our case, the S-duality can be understood to arise from the S-duality of the D5-NS5 system.

\mbox{}\par\nobreak
\noindent

\vspace{0.4cm}
\noindent{\large \bf Acknowledgements}
\vspace{0.3cm}

\noindent
We would like to thank Junya Yagi for discussions. This work is supported in part by the NUS FRC Tier 1 grant R-144-000-377-114.

\begin{thebibliography}{10}


\bibitem{Costello}
K.~Costello, {\it Supersymmetric gauge theory and the Yangian, ArXiV High-Energy Physics-Theory e-prints} (March, 2013) [\href{http://arxiv.org/abs/1303.2632}{{\tt arXiv:1303.2632}}]

\bibitem{CWY}
K.~Costello, E.~Witten, M.~Yamazaki, {\it Gauge Theory and Integrability, I, ArXiV High-Energy Physics-Theory e-prints} (September, 2017) [\href{http://arxiv.org/abs/1709.09993}{{\tt arXiv:1709.09993}}]


\bibitem{CWY2}
K.~Costello, E.~Witten, M.~Yamazaki, {\it Gauge Theory and Integrability, II, ArXiV High-Energy Physics-Theory e-prints} (February, 2018) [\href{http://arxiv.org/abs/1802.01579}{{\tt arXiv:1802.01579}}]

\bibitem{WittenKho}
E.~Witten, {\it Fivebranes and knots, ArXiV High-Energy Physics-Theory e-prints} (January, 2011) [\href{http://arxiv.org/abs/1101.3216}{{\tt arXiv:1101.3216}}] 

\bibitem{Witten}
E.~Witten, {\it Integrable Lattice Models From Gauge Theory, ArXiV High-Energy Physics-Theory e-prints} (November, 2016) [\href{http://arxiv.org/abs/1611.00592}{{\tt arXiv:1611.00592}}]

\bibitem{Geyer}
B.~Geyer, D.~M\"{u}lsch, {\it Higher-dimensional analogue of the Blau–Thompson model and NT= 8, D= 2 Hodge-type cohomological gauge theories},
{\em Nuclear Physics B} {\bf 662}
 (3) (2003) 531-553 [\href{http://arxiv.org/abs/hep-th/0211061}{{\tt arXiv:hep-th/0211061}}]

\bibitem{CJ}
C.~Cordova, D.~Jafferis, {\it Five-dimensional maximally supersymmetric Yang-Mills in supergravity backgrounds, ArXiV High-Energy Physics-Theory e-prints} (May, 2013) [\href{http://arxiv.org/abs/1305.2886}{{\tt arXiv:1305.2886}}] 

\bibitem{Bershadsky}
M.~Bershadsky, C.~Vafa, V.~Sadov, {\it D-branes and topological field theories},
{\em Nuclear Physics B} {\bf 463}
 (2-3) (1996) 420-434 [\href{http://arxiv.org/abs/hep-th/9511222}{{\tt arXiv:hep-th/9511222}}] 
 

\bibitem{CostelloYagi}
K.~Costello, J.~Yagi, {\it Unification of integrability in supersymmetric gauge theories, ArXiV High-Energy Physics-Theory e-prints} (October, 2018) [\href{http://arxiv.org/abs/1810.01970}{{\tt arXiv:1810.01970}}] 
 
\bibitem{Costello2}
K.~Costello, {\it M-theory in the Omega-background and 5-dimensional non-commutative gauge theory, ArXiV High-Energy Physics-Theory e-prints} (October, 2016) [\href{http://arxiv.org/abs/1610.04144}{{\tt arXiv:1610.04144}}]

\bibitem{Costello3}
K.~Costello, {\it Holography and Koszul duality: the example of the $ M2 $ brane, ArXiV High-Energy Physics-Theory e-prints} (May, 2017) [\href{http://arxiv.org/abs/1705.02500}{{\tt arXiv:1705.02500}}]

\bibitem{ACMV}
M.~Aganagic, K.~Costello, J.~McNamara, C.~Vafa, {\it Topological Chern-Simons/matter theories, ArXiV High-Energy Physics-Theory e-prints} (June, 2017) [\href{http://arxiv.org/abs/1706.09977}{{\tt arXiv:1706.09977}}]

 
\bibitem{TY} 
Y.~Terashima, M.~Yamazaki, {\it ${\text {SL}}\left ({2,\mathbb {R}}\right) $ Chern-Simons, Liouville, and Gauge Theory on Duality Walls}, {\em Journal of High Energy Physics} (2011) (8), 135 [\href{http://arxiv.org/abs/1103.5748}{{\tt arXiv:1103.5748}}] 
 
\bibitem{DimofteGukov}
 T.~Dimofte, S.~Gukov,  {\it Chern-Simons theory and S-duality}, {\em Journal of High Energy Physics} (2013) (5), 109 [\href{http://arxiv.org/abs/1106.4550}{{\tt arXiv:1106.4550}}] 
 
\end{thebibliography}


  
\vspace{0.4cm}

\appendix
\label{App:GammaMat}
\section{5d gamma matrices and spinor operations} 

Where necessary, we use the following explicit representation of the gamma matrices in five (Euclidean) dimensions
\begin{align}\label{eq:GammaMatrices}
\Gamma^1=\si^1\otimes\si^3, && \Gamma^2=\si^2\otimes\si^3, && \Gamma^3=\si^3\otimes\si^3 && \Gamma^4=\mathds{1}\otimes\si^1, && \Gamma^5=\mathds{1}\otimes\si^2, 
\end{align}
where the Pauli matrices $\{\si^1,\si^2,\si^3\}$ are
\begin{align}
\si^1=\left(
\begin{array}{cc}
0 & 1 \\
1 & 0
\end{array}\right),
&&
\si^2=\left(
\begin{array}{cc}
0 & -i \\
i & 0
\end{array}\right),
&&
\si^3=\left(
\begin{array}{cc}
1 & 0 \\
0 & -1
\end{array}\right).
\end{align}
These gamma matrices obey the Clifford algebra
\begin{equation} \label{gamma}
\left\{\Gamma_M,\Gamma_N\right\}=2g_{MN}\mathds{1}_{4\times 4}.
\end{equation}
In addition, we also use this set of gamma matrices for the R-symmetry group $SO(5)_R$.

The 
$SO(5)$ rotation/R-symmetry group spinor indices in this paper are raised and lowered using the two index antisymmetric tensor $\Omega$, i.e.,\footnote{We shall only write formulas corresponding to rotation group spinors in what follows; the corresponding formulas for R-symmetry group spinors can be obtained by replacing indices with hatted versions of themselves.}
\begin{align}
\rho_A=\rho^B\Omega_{BA}, && \rho^A=\Omega^{AB}\rho_B, \end{align}
where $\rho_A$ and $\rho^A$ correspond to the representation $\bf{4}$ and its dual representation $\bf{4}^{\vee}$.
Here, $\Omega$ is chosen to have the explicit form 
\begin{equation}
\Omega^{AB}=\eps^{\balpha\bbeta}\otimes B^{\bm\bn}=\begin{pmatrix} 0 & 1 \\ -1 & 0 \end{pmatrix}\otimes \begin{pmatrix} 0 & 1 \\ 1 & 0 \end{pmatrix}.
\end{equation}
Moreover, the two index antisymmetric tensor $\eps$ defined above can be used to raise and lower 
$SO(3)$ spinor indices, i.e.,
\begin{align}
\lambda_{\balpha}=\lambda^{\bbeta}\eps_{\bbeta\balpha}, && \lambda^{\balpha}=\eps^{\balpha\bbeta}\lambda_{\bbeta}. \end{align}
Note that this antisymmetric tensor acts on the Pauli matrices to give symmetric matrices, i.e.,  $(\si^{\alpha})_{\balpha}^{~~\bbeta}\eps_{\bbeta\bgamma}=(\si^{\alpha})_{\balpha\bgamma}$  and $\eps^{\balpha\bbeta}(\si^{\alpha})_{\bbeta}^{~~\bgamma}=(\si^{\alpha})^{\balpha\bgamma}$, where $(\si^{\alpha})_{\balpha\bgamma}=(\si^{\alpha})_{\bgamma\balpha}$ and $(\si^{\alpha})^{\balpha\bgamma}=(\si^{\alpha})^{\bgamma\balpha}$.

\end{document}